\documentclass[10pt,aps,prd,showpacs,twocolumn,sort&compress,floatfix,nofootinbib]{revtex4-1}

\usepackage{microtype}
\usepackage{upgreek}
\usepackage{amsmath,amssymb}
\usepackage{graphicx}

\usepackage[
    pdftitle={WW-RL-AWW-PS},
    pdfauthor={Dr. Thomas Hilger}
    bookmarks=true,
    pdfborder={0 0 0},
    bookmarksopen=true,
    unicode,
    hyperfootnotes=true,
    colorlinks=false,
    pagebackref=false,
    citecolor=blue,
    linkcolor=blue,
    urlcolor=blue,
    breaklinks,
    pdfpagelabels,
    ]{hyperref}
\usepackage[numbered]{bookmark}
\usepackage[all]{hypcap}

\usepackage[
    toc,
    acronym,
    nomain,
    style=long,
    nonumberlist,
    nowarn,
    ]{glossaries}

\makeglossaries
\glsdisablehyper
\newacronym{DS}{DS}{Dyson-Schwinger}
\newacronym{BS}{BS}{Bethe-Salpeter}
\newacronym{MT}{MT}{Maris-Tandy}
\newacronym{AWW}{AWW}{Alkofer-Watson-Weigel}
\newacronym{NG}{NG}{Nambu-Goldstone}
\newacronym{WW}{WW}{Wigner-Weyl}
\newacronym{SM}{SM}{Standard Model}
\newacronym{DSE}{DSE}{Dyson-Schwinger equation}
\newacronym{BSE}{BSE}{Bethe-Salpeter equation}
\newacronym{DSBS}{DSBS}{Dyson-Schwinger--Bethe-Salpeter}
\newacronym{BSA}{BSA}{Bethe-Salpeter amplitude}
\newacronym{BSM}{BSM}{Bethe-Salpeter matrix}
\newacronym{BSV}{BSV}{Bethe-Salpeter vertex}
\newacronym{QED}{QED}{quantum electrodynamics}
\newacronym{QCD}{QCD}{quantum chromodynamics}
\newacronym{QCDg}{QCD}{Quantenchromodynamik}
\newacronym{DCSB}{D$\upchi$SB}{dynamical chiral symmetry breaking}
\newacronym{VMD}{VMD}{vector meson dominance}
\newacronym{FAIR}{FAIR}{Facility for Antiproton and Ion Research}
\newacronym{HIC_for_FAIR}{HIC for FAIR}{Helmholtz International Center for FAIR}
\newacronym{WASAatCOSY}{WASA at COSY}{Wide Angle Shower Apparatus at Cooler Synchrotron}
\newacronym{WASA}{WASA}{Wide Angle Shower Apparatus}
\newacronym{COSY}{COSY}{Cooler Synchrotron}
\newacronym{MAMI}{MAMI}{\emph{Mainzer Mikrotron}}
\newacronym{LHC}{LHC}{Large Hadron Collider}
\newacronym{LHCb}{LHC{\scshape b}}{Large Hadron Collider beauty experiment}
\newacronym{KLOE}{KLOE}{K Long Experiment}
\newacronym{DAFNE}{DA$\Phi$NE}{Double Annular $\Phi$ Factory for Nice Experiments}
\newacronym{DESY}{DESY}{\emph{Deutsches Elektronen-Synchrotron}}
\newacronym{JLab}{JLab}{Thomas Jefferson National Accelerator Facility}
\newacronym{QSR}{QSR}{QCD sum rule}
\newacronym{OPE}{OPE}{operator product expansion}
\newacronym{JINR}{JINR}{Joint Institute for Nuclear Research}
\newacronym{HZDR}{HZDR}{Helmholtz-Zentrum Dresden-Rossendorf}
\newacronym{ANL}{ANL}{Argonne National Laboratory}
\newacronym{PANDA}{PANDA}{facility for anti-Proton ANnihilation at DArmstadt}
\newacronym{EFT}{EFT}{effective field theory}
\newacronym{VEV}{VEV}{vacuum expectation value}
\newacronym{KFU}{KFU}{Karl-Franzens University}
\newacronym{lQCD}{{\scshape l}QCD}{lattice-regularized QCD}
\newacronym{WTI}{WTI}{Ward-Takahashi identity}
\newacronym{AVWTI}{{\scshape av}WTI}{axialvector Ward-Takahashi identity}
\newacronym{RL}{RL}{rainbow ladder}
\newacronym{TUD}{TUD}{Technical University of Dresden}
\newacronym{UU}{UU}{Uppsala University}
\newacronym{NJL}{NJL}{Nambu--Jona-Lasinio model}
\newacronym{QFT}{QFT}{quantum field theory}
\newacronym{BRST}{BRST}{Becchi-Rouet-Stora-Tyutin}
\newacronym{FAKT}{FAKT}{\emph{``Fachausschuss Kern- und Teilchenphysik''}}
\newacronym{OPG}{{\"O}PG}{\emph{``{\"O}sterreichische Physikalische Gesellschaft''}}
\newacronym{APS}{{\"O}PG}{Austrian Physical Society}
\newacronym{SLAC}{SLAC}{Stanford Linear Accelerator Center}
\newacronym{DPG}{DPG}{\emph{``Deutsche Physikalische Gesellschaft''}}
\newacronym{GPS}{DPG}{German Physical Society}
\newacronym{FWF}{FWF}{\emph{``Fonds zur F{\"o}rderung der wissenschaftlichen Forschung''}}
\newacronym{KSU}{KSU}{Kent State University}
\newacronym{GMOR}{GMOR}{Gell-Mann--Oakes--Renner}
\newacronym{QGV}{QGV}{quark-gluon interaction vertex}
\newacronym{STI}{STI}{Slavnov-Taylor identity}
\newacronym{GU}{GU}{Giessen University}
\newacronym{GSI}{GSI}{\emph{Gesellschaft f{\"u}r Schwerionenforschung}}
\newacronym{ERA}{ERA}{European Research Area}
\newacronym{BRL}{BRL}{beyond-rainbow-ladder}
\newacronym{WP}{WP}{work package}
\newacronym{D}{D}{major deliverable}
\newacronym{BEPC}{BEPC}{Beijing Electron Positron Collider}
\newacronym{BES}{BES-III}{Bejing Spectrometer}
\newacronym{KEK}{KEK}{High Energy Accelerator Research Organization}

\newcommand{\ld}{: \!}
\newcommand{\rd}{\! :}

\newcommand{\Dp}[2]{\mathrm{d}^{#1\,} \! #2}

\newcommand{\mMeV}{\, {\text{M}}e{\text{V}}}

\newcommand{\mGeV}{ \, {\text{G}}e{\text{V}}}

\newlength{\fighght}
\AtBeginDocument{
\newboolean{draft}
\setboolean{draft}{true}
\newboolean{todo}
\setboolean{todo}{true}
\newcommand{\comment}[2]{\ifthenelse{\boolean{#1}}{\textcolor{blue}{\emph{\small #2}}}{}}
}

\begin{document}

\title{Poincar\'e covariant pseudoscalar and scalar meson spectroscopy in Wigner-Weyl phase}

\author{T.\,Hilger}

\email{thomas.hilger@uni-graz.at}

\affiliation{Institute of Physics, University of Graz, NAWI Graz, A-8010 Graz, Austria}

\date{\today}

\begin{abstract}
The coupled quark Dyson-Schwinger and meson Bethe-Salpeter equations in rainbow-ladder truncation for spin-0 mesons are solved in Wigner-Weyl phase in the chiral limit and beyond, retaining only the ultraviolet finite terms of the phenomenologically most successful Maris-Tandy interaction.
This allows to reveal and discuss the scalar and pseudoscalar meson masses in a chirally symmetric setting without additional medium effects.
Independent of the current-quark mass, the found solutions are spacelike, i.\,e.\ have negative squared masses.
The current-quark mass dependence of meson masses, leptonic decay constants and chiral condensate are illustrated in Wigner-Weyl phase.
\end{abstract}

\pacs{%
14.40.-n, 
%
%
%
%
12.38.Lg, 
%
%
11.10.St, 
%
%
11.30.Rd 
}

\maketitle

\section{Introduction}

The major mass fraction of visible matter in the universe originates in the non-perturbative momentum regime of \gls{QCD}.
There is consensus that \gls{DCSB} and/or confinement dynamically generate hadron masses which are several orders of magnitude larger than the current masses of the underlying valence constituents.
However, the interrelation of \gls{DCSB} and confinement is nowhere near understood and has recently been discussed in \cite{Greensite:2011zz,Pak:2015dxa,Glozman:2015ata,Biernat:2013fka} within a \gls{lQCD} approach.
While non-observable color charges and the non-degeneracy of chirality partners are clearly associated with confinement and \gls{DCSB},\footnote{These are the defining phenomena.} respectively, dynamically generated hadron masses are either attributed to the first or the second.
If confinement and \gls{DCSB} are tantamount and equivalent to each other, such a distinction is, of course, meaningless and an underlying mechanism causing both might be conceivable.
In any case, investigating the origin of mass is one way to approach this issue.

Therefore, future and existing large-scale experiments aim at studying the properties of visible matter under extreme conditions, i.\,e.\ large temperatures and/or densities, e.\,g.\ \cite{Friman:2011zz,Rapp:2011zz}, also in order to reveal if and how \gls{DCSB} and confinement are related to each other and if the mechanisms that are believed to generate the hadron properties are compatible with the physics observed there.
Here a phase transition or crossover towards the chirally symmetric and deconfined phase is anticipated.
In turn the properties of matter under extreme conditions, in particular under chiral symmetry restoration, are also theoretically widely discussed.
An interesting aspect of this endeavor is the question, what the properties of matter would be under the restoration of chiral symmetry isolated from density or temperature effects.

Poincar\'e covariant and symmetry preserving calculations of the hadron spectrum within a combined \gls{DS} equation and \gls{BS} equation approach \cite{Fischer:2006ub,Roberts:2007jh,Sanchis-Alepuz:2015tha} generate the crucial non-perturbative quark mass dressing which suffices to explain the non-exotic \cite{Hilger:2014nma,Maris:1999nt,Alkofer:2002bp,Maris:2005tt,Maris:2006ea,Krassnigg:2009zh,Krassnigg:2010mh,Blank:2011ha,Popovici:2014pha,Hilger:2014nma,Hilger:2015ora,Eichmann:2009qa,Sanchis-Alepuz:2014sca} and study the exotic \cite{Hilger:2015hka,Eichmann:2015cra,Burden:2002ps,Qin:2011xq} hadron spectrum.
Though this quark mass dressing is not an observable quantity, it is a valuable intuition building ingredient.
Here the \gls{AVWTI} ensures compliance with chiral symmetry and its dynamical breakdown at all energy scales \cite{Maris:1997tm,Maris:1997hd,Holl:2004fr}.
For a reliable hadron phenomenology it is thus a crucial constraint.
In view of this, investigating hadron properties in a world which is initially chirally symmetric within the \gls{DS}--\gls{BS} approach seems natural.
A chirally symmetric scenario without density and temperature might be considered as the simplest approximation to the complicated dynamics governing the QCD phase diagram, which still exhibits a phase transition.
In particular, such a scenario might show if restoration of chiral symmetry is sufficient for deconfinement.\footnote{Conversely, confinement would be sufficient for \gls{DCSB}.}
Deviations from the results and predictions of such a scenario are then attributed to medium effects.
Furthermore, the presented approach allows to study the effect of explicit chiral symmetry breaking and to compare it to \gls{DCSB} effects.

In the chiral quark mass limit, the pion would be massless in the \gls{NG} phase of chiral symmetry.
Its finite mass stems from the small explicit current-quark masses.
It is much smaller and clearly separated from the next heavier meson, the $\uprho$ meson.
While the mass of the latter can intuitively be understood by the large finite dynamically generated quark masses, the pion, with the same quark content, is so light due to its nature as a pseudo-Goldstone boson.
The latter must necessarily exist in the spectrum if chiral symmetry is spontaneously broken.
From this perspective it is by no means clear what the pion mass would be in the \gls{WW} phase of chiral symmetry, where neither significant dynamical quark masses are generated, nor the need for massless Goldstone bosons arises.

Experimentally, the pion dynamics and the pion mass in a \gls{WW} phase realization without additional medium effects is not known and, most probably, will never be known.
Nevertheless, in view of the unclear interrelation of confinement and \gls{DCSB}, in particular qualitative statements which are not interfered with additional medium effects, such as collisional broadening, are most enlightening and guide expectations.
Calculations within the Nambu-Jona-Lasinio model at nonzero temperature and baryon density \cite{Ratti:2004ra,Bernard:1987im} point to a (monotonically) increasing pion mass.
In \cite{Hilger:2010cn} the question about the $\uprho$ meson properties in a scenario which is chirally symmetric in vacuum has been posed and answered for the first time within \glspl{QSR}.

Similar investigations regarding the spectrum of spin-1 and 2 mesons, as well as nucleons, in such a scenario have recently been performed in \cite{Denissenya:2015woa,Denissenya:2015mqa,Glozman:2015qva}, where the authors observed a new SU(4) symmetry in the hadron spectrum.
Within these studies, no spin-0 mesons have been found in the meson spectrum.
In the same spirit one may employ the phenomenologically successful, symmetry preserving, coupled \gls{DS}--\gls{BS} approach to probe the effect of exclusive chiral symmetry restoration on hadronic properties as such and, in particular, disentangled from many-body effects.
In particular, this approach is genuinely Poincar\'e covariant and, thus, correctly reflects the related phenomena in the hadron spectrum, whatever these might be.
As the \gls{WW} spectrum is experimentally unknown and guidance is missing from this side, such a Poincar\'e covariant approach is well justified.
Early studies with the same scope have also been done in \cite{Bicudo:2006dn}, and references therein, for simplified confining models.
This is feasible because the coupled \gls{DS}--\gls{BS} approach naturally entails solutions in the \gls{WW} phase.
Note, that the \gls{WW} phase solution to the \gls{RL} truncated \gls{DS}-\gls{BS} equation approach is as valid and consistent as the \gls{NG} solution.
In this spirit, it has the advantage that it does not require a deformation of the theory contrary to current \gls{QSR} or \gls{lQCD} based approaches.
On the other hand, one is forced to employ a model interaction rather than a first principle QCD calculation.
Nevertheless, because the \gls{MT} model, on which the employed \gls{AWW} model \cite{Alkofer:2002bp} is based, is phenomenologically very successful with a solid and well-founded relation to QCD, enlightening results are found.

In section~\ref{sct:dse} and \ref{sct:bse} the quark \gls{DS} and meson \gls{BS} equations in \gls{RL} truncation together with the employed \gls{AWW} model interaction are presented.
Results are presented and discussed in section~\ref{sct:results} with conclusions in section \ref{sct:conclucions}.

\section{Quark DSE}
\label{sct:dse}

The employed \gls{DS} equation in rainbow truncation for the non-perturbative quark propagator reads
\begin{subequations}\label{eq:dse}
\begin{align}
    &S(p)^{-1} = Z_2 \, \left(i\gamma\cdot p + Z_4 \, m_q\right) + \Sigma(p) \,,
    \\
    &\Sigma(p) = C_\mathrm{F} \, \int^\Lambda_q\!\! \mathcal{G}\!\left((p-q)^2\right) \, D_{\mu\nu}^\mathrm{f}(p-q) \,\gamma_\mu \,S(q)\, \gamma_\nu \,,
\end{align}
\end{subequations}
where the Casimir color factor is given as $C_\mathrm{F} = {(N_\mathrm{c}^2-1)}/{2N_\mathrm{c}}$ and the number of color degrees of freedom being $N_\mathrm{c} = 3$.
$\Sigma(p)$ is called the quark self-energy or mass shift operator and $D_{\mu\nu}^\mathrm{f}(l) = \left(\delta_{\mu\nu} - {l_\mu l_\nu}/{l^2} \right)$ is the transversal projector part of the free gluon propagator in Landau gauge.
The model interaction $l^2 \mathcal{G}(l^2)$ algebraically replaces the gluon propagator dressing function and is intended to imitate the combined effects of omitted quark-gluon vertex terms, gluon propagator dressing function and running coupling.
$\int^\Lambda_q = \int^\Lambda \frac{\Dp{4}{q}}{(2\pi)^4}$ is a translationally invariant regularized integration measure with regularization scale $\Lambda$ \cite{Maris:1997tm}.
$Z_2$ and $Z_4$ are quark wave function and quark mass renormalization constants, whereas the quark gluon-vertex renormalization constant is absorbed in the model interaction $\mathcal{G}((p-q)^2)$.
The current-quark mass is denoted by $m_q$.

With the decompositions
\begin{equation}\label{eq:dressing}
\begin{split}
    S(p)^{-1} &= i\gamma\cdot p \; A(p^2)+B(p^2)
    = Z(p^2)\left(i\gamma\cdot p+M(p^2)\right)
    \\
    &= \left[ -i\gamma\cdot p \; \sigma_\mathrm{V}(p^2) + \sigma_\mathrm{S}(p^2) \right]^{-1} \,,
\end{split}
\end{equation}
Eq.\,\eqref{eq:dse} defines a system of inhomogeneous, non-linear, singular, coupled Fredholm integral equations of the second kind for the propagator dressing functions $A$ and $B$.
Depending upon details of the specific interaction, e.\,g.\ strength, multiple solutions exist.
Most commonly two different solution strategies are employed: fixed-point iteration and optimization algorithms.

Solutions in the chiral limit with non-zero chiral quark condensate are identified as solutions in \gls{NG} phase.
Those with zero chiral condensate are assumed to be solutions in \gls{WW} phase.%
\footnote{
Note, that in general a zero chiral condensate is necessary but not sufficient for chiral symmetry realizations.
See also the discussion in Sec.~\ref{sct:results}.
}
Furthermore, we assume that the transition from one solution in a certain phase of chiral symmetry to another solution in the same phase but for a different current-quark mass is continuous in the current-quark mass $m_q$.
This allows to identify solutions in \gls{WW} phase beyond the chiral limit.
Finally it is assumed that the gluon and quark-gluon dynamics is not affected by a transition to the chirally symmetric phase or that the corresponding corrections are at least sufficiently small to give reasonable qualitative results when neglected.
Clearly, due to the coupling of gluon propagator and quark-gluon-vertex to the quark propagator, see e.\,g.\ \cite{Alkofer:2014taa,Williams:2014iea} or \cite{Gomez-Rocha:2015qga,Gomez-Rocha:2014vsa} for an intuitive model, this is an approximation.
Up to now, the gluon propagator in \gls{WW} phase is unknown.

The phenomenologically most successful model within the \gls{RL} truncated \gls{DS}--\gls{BS} approach to the meson spectrum is the \gls{MT} model developed in \cite{Maris:1999nt} with the most recent applications to a comprehensive meson phenomenology in \cite{Hilger:2015ora,Hilger:2015hka,Hilger:2014nma,Popovici:2014pha}.
It consists of an infrared and an ultraviolet part.
The latter is crucial to ensure the proper perturbative limit of the running coupling in QCD.
However, it is of minor qualitative importance for meson spectroscopy \cite{Alkofer:2002bp,Jain:1993qh}.
In particular it can be neglected when mainly qualitative aspects are investigated, as in the scope of the current investigation.

\begin{figure*}
\centering
\includegraphics[width=.49\textwidth]{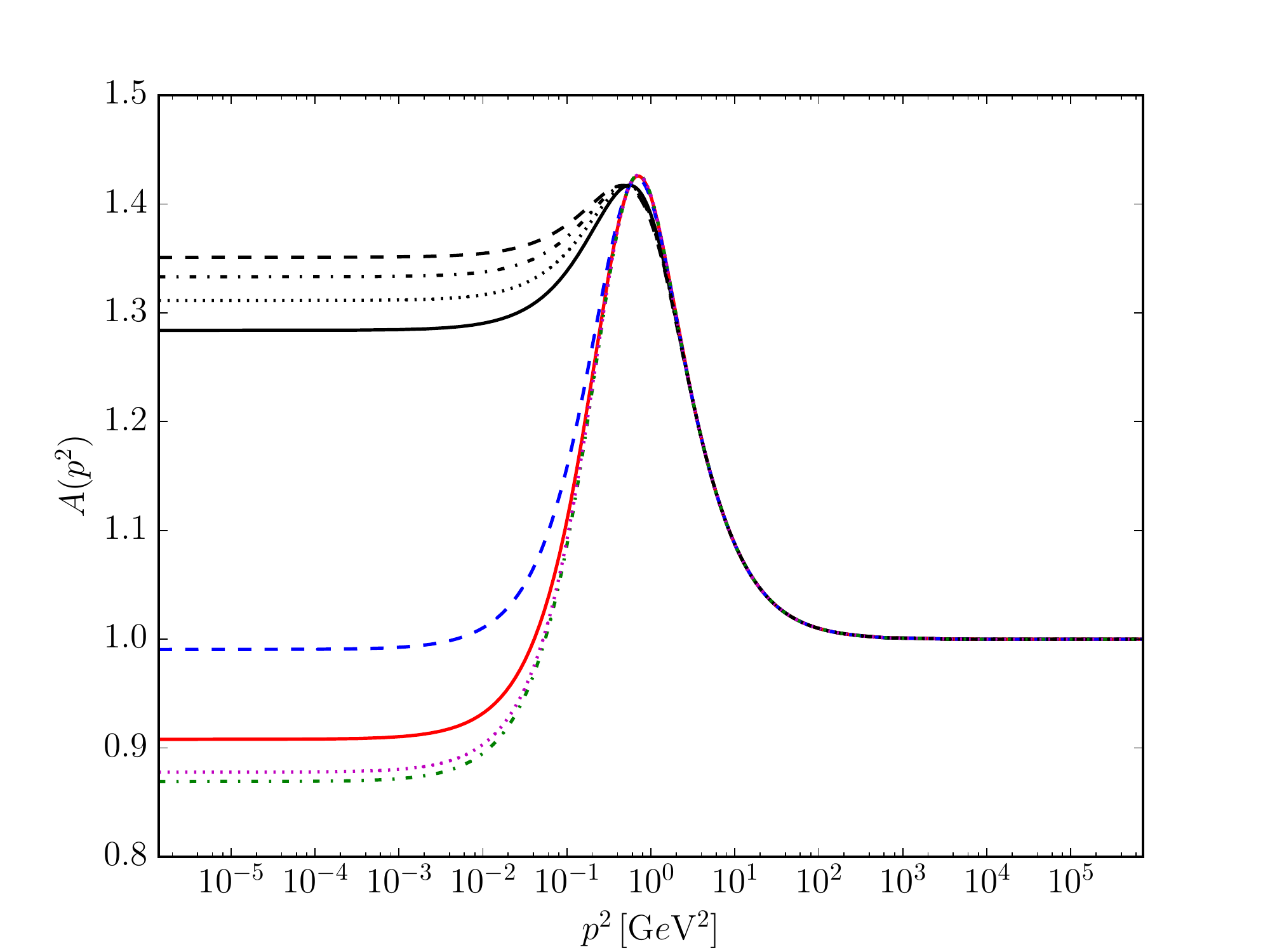}
\includegraphics[width=.49\textwidth]{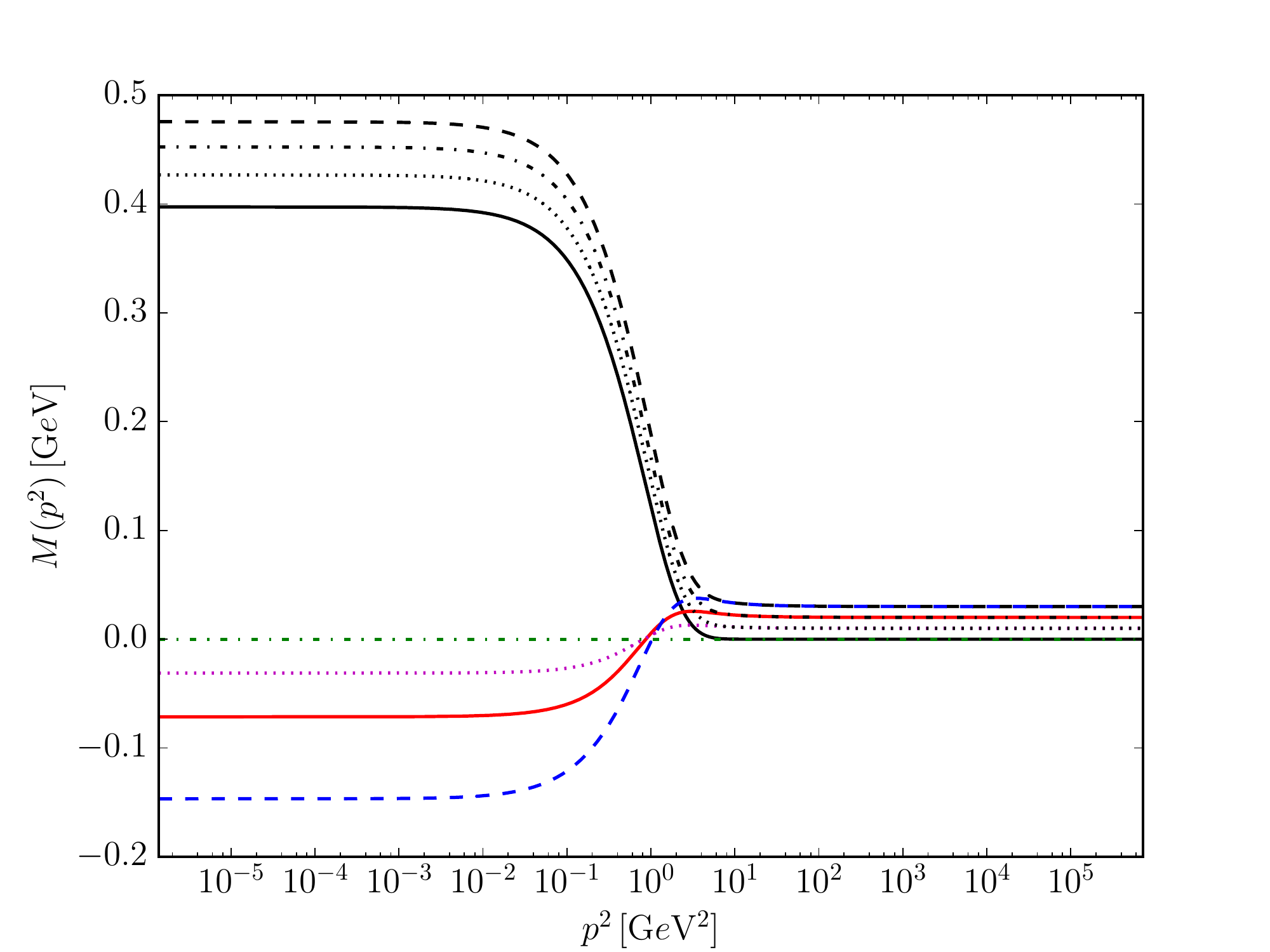}
\caption[Solutions of the DSE in the Nambu-Goldstone and Wigner-Weyl phase]{
Propagator functions $A(p^2)$ (left panel) and $M(p^2)$ (right panel) along the real axis obtained from Newton-Krylov optimization for different initial functions $B_0(p^2)$ at $\omega = 0.5$\,G$e$V and $D=16.0$\,G$e$V$^2$ in \gls{WW} phase for $m_q = 0$ (solid red), $m_q = 10$\,M$e$V (dashed blue), $m_q = 20$\,M$e$V (dashed-dotted green) and $m_q = 30$\,M$e$V (dotted magenta).
Solutions in \gls{NG} phase are depicted in black with line styles which correspond to the \gls{WW} phase.
}
\label{fig:DSEWW}
\end{figure*}

The \gls{AWW} parametrization \cite{Alkofer:2002bp} of the interaction reads
\begin{equation}\label{eq:AWW}
    \mathcal{G}(q^2) = 4 \pi^2 D \frac{q^2}{\omega^2} e^{-\frac{q^2}{\omega^2}} \, .
\end{equation}
It is ultraviolet finite, all renormalization constants are equal to one and the limit $\Lambda \to \infty$ can be taken initially.
In particular, it obeys a chirally symmetric solution of Eq.\,\eqref{eq:dse} if $m_q = 0$, which can easily be obtained by employing $B_0(p) = 0$ as initial function for a fixed-point iteration.
It features $M(p) = 0$ with $A(p) \neq 0$; cf.\ Fig.\,\ref{fig:DSEWW} for a comparison to the chiral \gls{NG} phase solution and the employed parameters.
Both dressing functions have the same asymptotic behavior in the \gls{WW} phase as in the \gls{NG} phase.
While $M(p)$ in \gls{WW} phase significantly deviates from its \gls{NG} phase, $A(p)$ differs only below $\approx 1$\,G$e$V.

Among others, the Newton-Krylov root finding method \cite{Knoll2004357}, which is suitable for large scale optimizations, can be used to find solutions in \gls{WW} phase beyond the chiral limit as well \cite{Bloch:1995dd,Williams:2007zzh}.
Figure \ref{fig:DSEWW} depicts the \gls{WW} solution for different bare current-quark masses $m_q$ up to a critical mass $m_q^\mathrm{cr} = 31$\,M$e$V, above which no solution in \gls{WW} phase has been found for the employed set of parameters in the present setup.
The quark mass functions $M(p)$ resemble roots at $\approx 1$\,G$e$V and deviate significantly from each other.
Again, the propagator functions $A(p)$ only deviate significantly below $\approx 1$\,G$e$V from each other.
The characteristic excess of this function at $\approx 1$\,G$e$V remains unaffected.
Consequently, even in the chiral limit and \gls{WW} phase a simple constituent quark picture or the modeling of quark bound states by virtue of free quark propagators is not applicable or justified in the light quark sector.
Finally, both figures resemble the relevance of the $1$\,G$e$V scale for \gls{DCSB} and its restoration.
However, the rigorous interrelation of the characteristic \gls{DCSB} scale and the critical current-quark mass $m_q^\mathrm{cr}$ within this model remains unknown.

\section{Meson BSE}
\label{sct:bse}

In order to respect the fundamental symmetries of the underlying interaction, the kernels of \gls{DS} and \gls{BS} equation must satisfy the \gls{AVWTI}.
This can be shown to be true for the \gls{RL} truncated \gls{DS}--\gls{BS} approach.
The \gls{RL} truncated homogeneous meson \gls{BS} equation reads
\begin{multline}\label{eq:BSE}
    \Gamma(p;P) = -C_\mathrm{F}\, Z_2^2\!\!\int^\Lambda_q\!\!\!\!\mathcal{G}((p-q)^2)\; D_{\mu\nu}^\mathrm{f}(p-q)
        \\
        \times \gamma_\mu \; S_1(q_+) \Gamma(q;P) S_2(q_-) \;\gamma_\nu \,,
\end{multline}
with $q_{\pm} = q \pm [1/2 \pm (\eta - 1/2)] P$ and $\Gamma(p;P)$ is the \gls{BSA}.
In what follows, the momentum partitioning parameter is set to $\eta=1/2$.

The solution strategy and covariant basis follow Refs.\ \cite{UweHilger:2012uua,Dorkin:2013rsa,Dorkin:2010ut,Blank:2010bp}.
The \gls{BS} amplitude is expanded in a finite set of covariants $T_n(p,P,\gamma)$, which specify the quantum numbers $J$ and $P$, according to
\begin{equation}
    \Gamma(p;P) = \sum_{n} \Gamma_{n}\!\!\left(p^2,\cos\sphericalangle(P,p);P^2\right) \, T_n(p,P,\gamma) \; ,
\end{equation}
with $\Gamma_{n}\!\!\left(p^2,\cos\sphericalangle(P,p);P^2\right)$ being the partial amplitudes.
Canonical normalization, leptonic decay constants, residual and (generalized) \gls{GMOR} relation are evaluated according to \cite{Maris:1999nt}.

Having the propagator functions in \gls{WW} phase for zero current-quark mass at disposal, one can solve the \gls{BS} equation in \gls{WW} phase and obtain an explicit value of the bound state mass.
As in \cite{Bicudo:2006dn}, the lowest bound state mass where the $J=0$-\gls{BS} equation \eqref{eq:BSE} in the chiral limit and the parameters of Fig.\,\ref{fig:DSEWW} can be solved is found at spacelike masses $M^2 = -0.1172 \mGeV^2 = -(342.1 \mMeV)^2$, which are called tachyonic solutions.
For timelike bound state momenta, the \gls{BS} equation integration domains of the quark propagators extend to the complex plane and the analytical structure must be accounted for \cite{Dorkin:2013rsa,Dorkin:2014lxa}.
For spacelike momenta, the integration domain is limited to the positive real axis where no non-analytical behaviour has been observed in \gls{WW} phase and solution of the coupled \gls{DS}--\gls{BS} system is straight forward.

As argued in \cite{Jain:2007zz} the \gls{WW} (tachyonic) solution corresponds to a maximum of the effective action.
Therefore, the squared mass must be negative and signals the instability of the chirally symmetric ground state.
An arbitrary small disturbance drives the system from the \gls{WW} realization to the \gls{NG} realization.
Conversely, if a stable chirally symmetric phase is to be realized at high densities and/or temperatures, the here neglected medium effects must, thus, eliminate all tachyonic solutions.

For completeness it is noted that the result for the \gls{NG} pseudoscalar boundstate mass is $M_\uppi=137$\,M$e$V with $f_\uppi = 94.1$\,M$e$V and the \gls{NG} scalar boundstate mass is $M_\upsigma = 645$\,M$e$V for $m_q = 5$\,M$e$V and the parameters as in Fig.\,\ref{fig:DSEWW}.

\section{Results and discussion}
\label{sct:results}

The analytic properties of the quark propagator may be analysed as in \cite{Dorkin:2013rsa} by Cauchy's argument principle or utilizing a Newton-Krylov root finding method.
In \gls{NG} phase, the quark propagator has a tower of complex conjugated poles off the real axis \cite{Dorkin:2013rsa}.
In \gls{WW} phase, the pole which is closest to the origin, i.\,e.\ the first relevant pole for timelike boundstates, can be found at $q^2 \approx -0.225 \cdot 10^{-3}$\,G$e$V$^2$, i.\,e.\ on the real axis and rather close to the origin.
It corresponds to a maximal accessible quarkonia boundstate mass of $M \approx 30$\,M$e$V if the pole has to be kept outside of the integration domain in Eq.\,\eqref{eq:BSE}.
Within the momentum region $-1\,\mathrm{G}e\mathrm{V}^2 \leq \mathrm{Re}\,p^2 \leq 0$, $\left| \mathrm{Im}\,p^2 \right| \leq 1\,\mathrm{G}e\mathrm{V}^2$ no complex conjugated poles (off the real axis) have been found.
Furthermore, apart from $\sigma_\mathrm{V}$, all propagator functions for $m_q = 5$\,M$e$V, cf.\ Eq.\,\eqref{eq:dressing}, have inflection points below $3$\,G$e$V$^2$ in \gls{WW} phase.
Such inflection points have been argued in, e.\,g.\ \cite{Roberts:2007ji}, to be sufficient for confinement.
Clearly, for $m_q = 0$ the propagator functions $B(p^2)$, $M(p^2)$ and $\sigma_S(p^2)$ have no inflection points in \gls{WW} phase.

The fact that the chiral limit \gls{WW} solution is compatible with a realization of chiral symmetry, i.\,e.\ Eq.\,\eqref{eq:BSE} gives identical \gls{BS} matrices and \gls{BS} amplitudes for chiral partner mesons and degenerate mass spectra, can be seen in a two-fold way.

First, the chiral condensate, which is given within the employed model \eqref{eq:AWW} for the gluon propagator by \cite{Zong:2003kf,Langfeld:2003ye}
\begin{equation}\label{eq:cond}
    \langle \ld \bar{q} q \rd \rangle = - \frac{3}{4\pi^2} \int_0^\infty \! \Dp{2}{l} \, l^2 \sigma_\mathrm{S}(l^2)
    \: ,
\end{equation}
is zero, because $B(p) = 0$.
In \gls{NG} phase a value of $\langle\ld\bar{q}q\rd\rangle = \left(-251\mMeV\right)^3$ is obtained for the employed set of parameters, which is in agreement with the (traditional) \gls{GMOR} relation
\begin{equation} \label{eq:GOR}
    f_\uppi^2 M_\uppi^2 = - 2 m_q \langle \ld \bar q q \rd \rangle
    \: .
\end{equation}
The leptonic decay constant of the pion within the \gls{AWW} model is
\begin{equation}
    \frac{f_\uppi}{3} = \left. \int_q \! \frac{\mathrm{Tr}[\gamma_5 \gamma\!\cdot\! P \; \chi^{0^-}\!\!(q;P) ]}{\sqrt{2}P^2} \right|_{P^2 = - M_\uppi^2}
    \,,
\end{equation}
with
$$\chi^{0^-}\!\!(q;P) \equiv S_1(q_+) \, \Gamma^{0^-}\!\!(q;P) \, S_2(q_-)$$
and $\Gamma^{0^-}\!\!(q;P)$ the pion's \gls{BSA}.
Equation \eqref{eq:GOR} can be extended to the above mentioned generalized \gls{GMOR} relation \cite{Maris:1997tm,Maris:1997hd,Holl:2004fr}
\begin{equation} \label{eq:GMOR}
    f_\uppi M_\uppi^2 = 2 m_q r_\uppi \,,
\end{equation}
where the residue of the pion mass pole in the pseudoscalar vertex of the \gls{AWW} model is given by \cite{Maris:1997tm,Maris:1997hd}
\begin{equation}
    \mathrm{i} r_\uppi = \left. \int_q \frac{\mathrm{Tr} [\gamma_5 \; \chi^{0^-}\!\!(q;P) ]}{\sqrt{2}} \right|_{P^2 = - M_\uppi^2} \,.
\end{equation}
Equation \eqref{eq:GMOR} is valid for all current-quark masses and pseudoscalar states.
Thus, it provides a natural definition of the chiral condensate which is valid beyond the chiral limit \cite{Holl:2004fr,Roberts:2012sv}:
\begin{equation} \label{eq:qq}
    \langle \ld \bar q q \rd \rangle \equiv - f_\uppi \: r_\uppi \,.
\end{equation}

As the chiral condensate transforms non-trivially under the chiral transformation group, vanishing of the chiral condensate is a necessary requirement for restoration of the symmetry.
However, strictly speaking the vanishing of an order parameter, which is qualified as such solely by means of its transformation properties, is merely a necessary but not sufficient requirement for the realization of a symmetry.
The realization of other symmetries or complicated medium effects may lead to vanishing condensates as well, which has been discussed in some detail for four-quark condensates in \cite{Thomas:2007es,Thomas:2007gx}.
A rather drastic example has recently been discussed in \cite{Buchheim:2015xka,Buchheim:2014rpa}.
In this spirit, determination of the phase transition temperature and/or density by virtue of the vanishing chiral condensate \eqref{eq:cond} only gives a lower bound as all non-trivially transforming condensates must be zero in the chirally symmetric phase.
In view of this, an accidental simultaneous vanishing of all such condensates with increasing temperature and/or density would indicate a systematic interrelation among these condensates which is up to now not known.
Within the context of open flavor chiral partner \glspl{QSR} \cite{Hilger:2009kn,Hilger:2010zb,Hilger:2012db,Hilger:2011cq}, considering a zero chiral condensate as a sufficient condition for chiral symmetry restoration, in the sense that no non-trivially transforming condensate points to a larger restoration temperature/density, corresponds to the claim that the lowest spectral moment of chiral partner spectra is the last one to vanish w.\,r.\,t.\ increasing temperature/density.%
\footnote{As the chiral condensate is the non-trivially transforming condensate with the lowest mass dimension, such a scenario seems indeed natural or at least tempting.
However, scenarios with \gls{DCSB} and zero chiral condensate have been discussed some time ago in \cite{Stern:1997ri,Stern:1998dy,Kogan:1998zc} and recently again in \cite{Kanazawa:2015kca}.}
There is no proof of such a claim up to now.

Second, it can be seen by degeneracy of the solutions to the \gls{BS} equation for chiral partners simply by the fact that the quark mass functions are zero.
In \gls{RL} truncation it can be shown, that the \gls{BS} equation integral kernels for scalar and pseudoscalar mesons only differ by terms proportional to quark mass functions $M(p)$ \cite{UweHilger:2012uua}.
Since the chiral limit \gls{BS} equations for scalar and pseudoscalar mesons are, therefore, identical in \gls{WW} phase, the \gls{BS} amplitudes and, hence, any observables are, too.
Strictly speaking degeneracy of the spin-0, or any other subset of the meson spectrum, alone is again, as in the case of non-trivially transforming condensates, not a sufficient but only a necessary requirement for chiral symmetry.
All chiral partner spectral densities must be degenerate.
However, as can be seen from chiral partner \glspl{QSR} \cite{Hilger:2011cq,Kapusta:1993hq} the spectra of chiral partners can only be degenerate if a complete hierarchy of non-trivially transforming condensates is zero.
In particular it has been explicitly demonstrated that if, e.\,g., scalar and pseudoscalar open flavor mesons are degenerate within the scope of \glspl{QSR}, the same holds true for vector and axialvector open flavor mesons \cite{Hilger:2011cq}.

\begin{figure*}
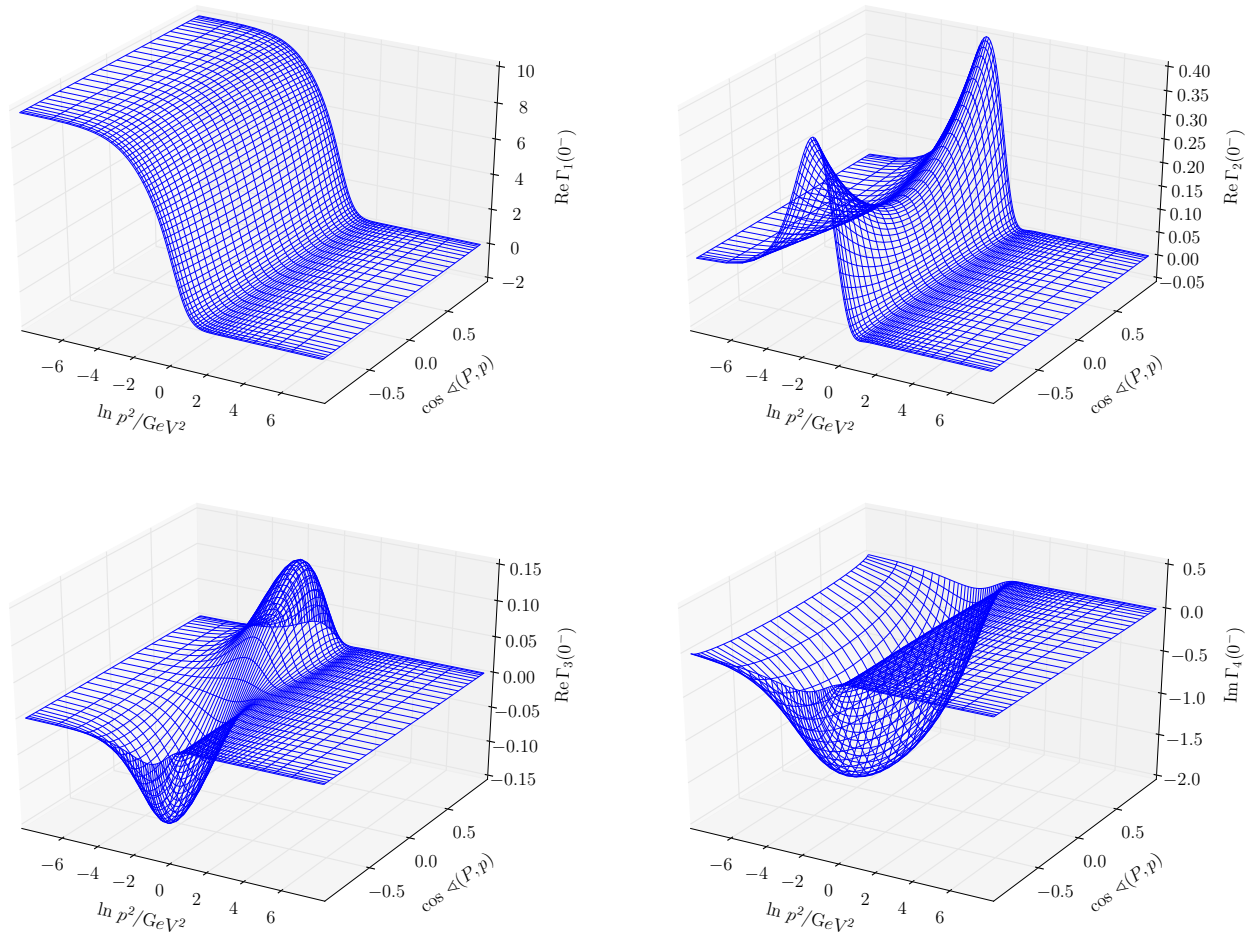
%
\centering
\includegraphics[width=.49\textwidth]{{{ReBSA0-+_SEARCH_AWW_WW_o0.5_D1.0_EX00_x32x32x64_graz4_M0.00+0.10j_0.00+0.37j_-0x5442803998cda6bd-1}}}
\includegraphics[width=.49\textwidth]{{{ReBSA0-+_SEARCH_AWW_WW_o0.5_D1.0_EX00_x32x32x64_graz4_M0.00+0.10j_0.00+0.37j_-0x5442803998cda6bd-2}}}%
\\
\includegraphics[width=.49\textwidth]{{{ReBSA0-+_SEARCH_AWW_WW_o0.5_D1.0_EX00_x32x32x64_graz4_M0.00+0.10j_0.00+0.37j_-0x5442803998cda6bd-3}}}
\includegraphics[width=.49\textwidth]{{{ImBSA0-+_SEARCH_AWW_WW_o0.5_D1.0_EX00_x32x32x64_graz4_M0.00+0.10j_0.00+0.37j_-0x5442803998cda6bd-4}}}%
\caption[Pseudoscalar BS amplitude in WW phase.]{
Non-zero parts of the pseudoscalar \gls{BS} partial amplitudes in \gls{WW} phase for $m_q = 5$\,M$e$V, $\omega = 0.5$\,G$e$V and $D=16.0$\,G$e$V$^2$.
All other parts are $\lessapprox 10^{-14}$.
}%
\label{fig:BSA_WW_P}%
\end{figure*}

\begin{figure*}%
\centering
\includegraphics[width=.49\textwidth]{{{ImBSA0++_SEARCH_AWW_WW_o0.5_D1.0_EX00_x32x32x64_graz4_M0.00+0.22j_0.00+0.39j_-0x1e8196eaa62bdfe8-1}}}
\includegraphics[width=.49\textwidth]{{{ReBSA0++_SEARCH_AWW_WW_o0.5_D1.0_EX00_x32x32x64_graz4_M0.00+0.22j_0.00+0.39j_-0x1e8196eaa62bdfe8-2}}}%
\\
\includegraphics[width=.49\textwidth]{{{ReBSA0++_SEARCH_AWW_WW_o0.5_D1.0_EX00_x32x32x64_graz4_M0.00+0.22j_0.00+0.39j_-0x1e8196eaa62bdfe8-3}}}
\includegraphics[width=.49\textwidth]{{{ReBSA0++_SEARCH_AWW_WW_o0.5_D1.0_EX00_x32x32x64_graz4_M0.00+0.22j_0.00+0.39j_-0x1e8196eaa62bdfe8-4}}}%
\caption[Scalar BS amplitude in WW phase.]{
Non-zero parts of the scalar \gls{BS} partial amplitudes in \gls{WW} phase for $m_q = 5$\,M$e$V, $\omega = 0.5$\,G$e$V and $D=16.0$\,G$e$V$^2$.
All other parts are $\lessapprox 10^{-15}$.
}%
\label{fig:BSA_WW_S}%
\end{figure*}

In Figs.\,\ref{fig:BSA_WW_P} and \ref{fig:BSA_WW_S} the non-zero parts of the complex canonically normalized partial amplitudes for scalar and pseudoscalar mesons in \gls{WW} phase at $m_q = 5$\,M$e$V are exhibited.
The major non-zero amplitudes, $\Gamma_1$, are very similar in both channels.
The other amplitudes drastically differ.
Note, that imaginary and real part of some partial amplitudes interchange their roles, i.\,e.\ what is zero in one channel is non-zero in the other.
This is related to the particular choice of covariants and $P^2$ being spacelike.
For comparison, the non-zero parts of the complex canonically normalized partial amplitudes in \gls{NG} phase at $m_q = 5$\,M$e$V for scalar and pseudoscalar mesons are exhibited in Figs.\,\ref{fig:BSA_NG_P} and \ref{fig:BSA_NG_S}.
The changing pattern is similar but not identical.
For example, imaginary and real parts of all partial amplitudes interchange their roles.
Turning from \gls{NG} to \gls{WW} phase in the scalar channel, all but the forth partial amplitude switch from zero real to zero imaginary part and vice versa.
Contrary, in the pseudoscalar channel, only the first partial amplitude switches.
Furthermore, up to a sign change in $\Gamma_4^{0^-}$, all amplitudes show the same qualitative behavior, in particular the same symmetries.
In the scalar channel, there is no sign change when passing over from one phase to the other.

In \cite{Maris:1997tm} it has been shown that
\begin{equation}
    f_\uppi \left. \Gamma_{1}^{0^-}(p^2,\cos\sphericalangle(P,p);P^2) \right|_{P^2=0} = 2 B(p^2)
\end{equation}
follows from the chiral limit \gls{AVWTI}.
As $B(p^2)$ is zero in the chiral limit \gls{WW} phase solution and the partial amplitude $\Gamma_{1}^{0^-}$, as well as all other amplitudes, are not, $f_\uppi = 0$ must hold in order not to violate the \gls{AVWTI}.
Indeed, for $m_q =0$ one has $f_\uppi \lessapprox 10^{-14}$\,G$e$V within this study, which numerically confirms the above conclusion.

\begin{figure*}%
\centering
\includegraphics[width=.49\textwidth]{{{ImBSA0-+_SEARCH_AWW_NG_o0.5_D1.0_EX00_x32x32x64_graz4_M0.10_0.17_0x527c779145ad169f-1}}}
\includegraphics[width=.49\textwidth]{{{ReBSA0-+_SEARCH_AWW_NG_o0.5_D1.0_EX00_x32x32x64_graz4_M0.10_0.17_0x527c779145ad169f-2}}}%
\\
\includegraphics[width=.49\textwidth]{{{ReBSA0-+_SEARCH_AWW_NG_o0.5_D1.0_EX00_x32x32x64_graz4_M0.10_0.17_0x527c779145ad169f-3}}}
\includegraphics[width=.49\textwidth]{{{ImBSA0-+_SEARCH_AWW_NG_o0.5_D1.0_EX00_x32x32x64_graz4_M0.10_0.17_0x527c779145ad169f-4}}}%
\caption[Pseudoscalar BS amplitude in NG phase.]{
Non-zero parts of the pseudoscalar \gls{BS} partial amplitudes in \gls{NG} phase (pion) for $m_q = 5$\,M$e$V, $\omega = 0.5$\,G$e$V and $D=16.0$\,G$e$V$^2$.
All other parts are $\lessapprox 10^{-14}$.
}%
\label{fig:BSA_NG_P}%
\end{figure*}

\begin{figure*}%
\centering
\includegraphics[width=.49\textwidth]{{{ReBSA0++_SEARCH_AWW_NG_o0.5_D1.0_EX00_x32x32x64_graz4_M0.03_0.71_0x6974c4cf88f3d50d-1}}}
\includegraphics[width=.49\textwidth]{{{ImBSA0++_SEARCH_AWW_NG_o0.5_D1.0_EX00_x32x32x64_graz4_M0.03_0.71_0x6974c4cf88f3d50d-2}}}%
\\
\includegraphics[width=.49\textwidth]{{{ImBSA0++_SEARCH_AWW_NG_o0.5_D1.0_EX00_x32x32x64_graz4_M0.03_0.71_0x6974c4cf88f3d50d-3}}}
\includegraphics[width=.49\textwidth]{{{ReBSA0++_SEARCH_AWW_NG_o0.5_D1.0_EX00_x32x32x64_graz4_M0.03_0.71_0x6974c4cf88f3d50d-4}}}%
\caption[Scalar BS amplitude in NG phase.]{
Non-zero parts of the scalar \gls{BS} partial amplitudes in \gls{NG} phase (sigma) for $m_q = 5$\,M$e$V, $\omega = 0.5$\,G$e$V and $D=16.0$\,G$e$V$^2$.
All other parts are $\lessapprox 10^{-14}$.
}%
\label{fig:BSA_NG_S}%
\end{figure*}

In \cite{Williams:2006vva,Williams:2007ef,Williams:2007ey,Williams:2007zzh} the \gls{WW} solution and so-called noded solutions\footnote{%
Solutions of the \gls{DS} equation in \gls{NG} phase do not have roots along the positive real axis.
Solutions in \gls{WW} phase have one root (node).
Other solutions have more than one node and are, therefore, dubbed noded solutions.
In analogy to vibrating strings, they are sometimes referred to as excited solutions \cite{Martin:2006qd}.}
to the \gls{DS} equation have been used to discuss the chiral condensate beyond the chiral limit.
Linear combinations of the quark propagators have been introduced, which all generate identical condensates in the chiral limit.
However, due to the non-linearity of the \gls{DS} equation, a linear combination of solutions cannot fulfill the corresponding \gls{DS} equation and, therefore, does not represent a self-consistent solution to the given \gls{DS} equation.

\begin{figure}%
\centering
\includegraphics[width=.45\textwidth]{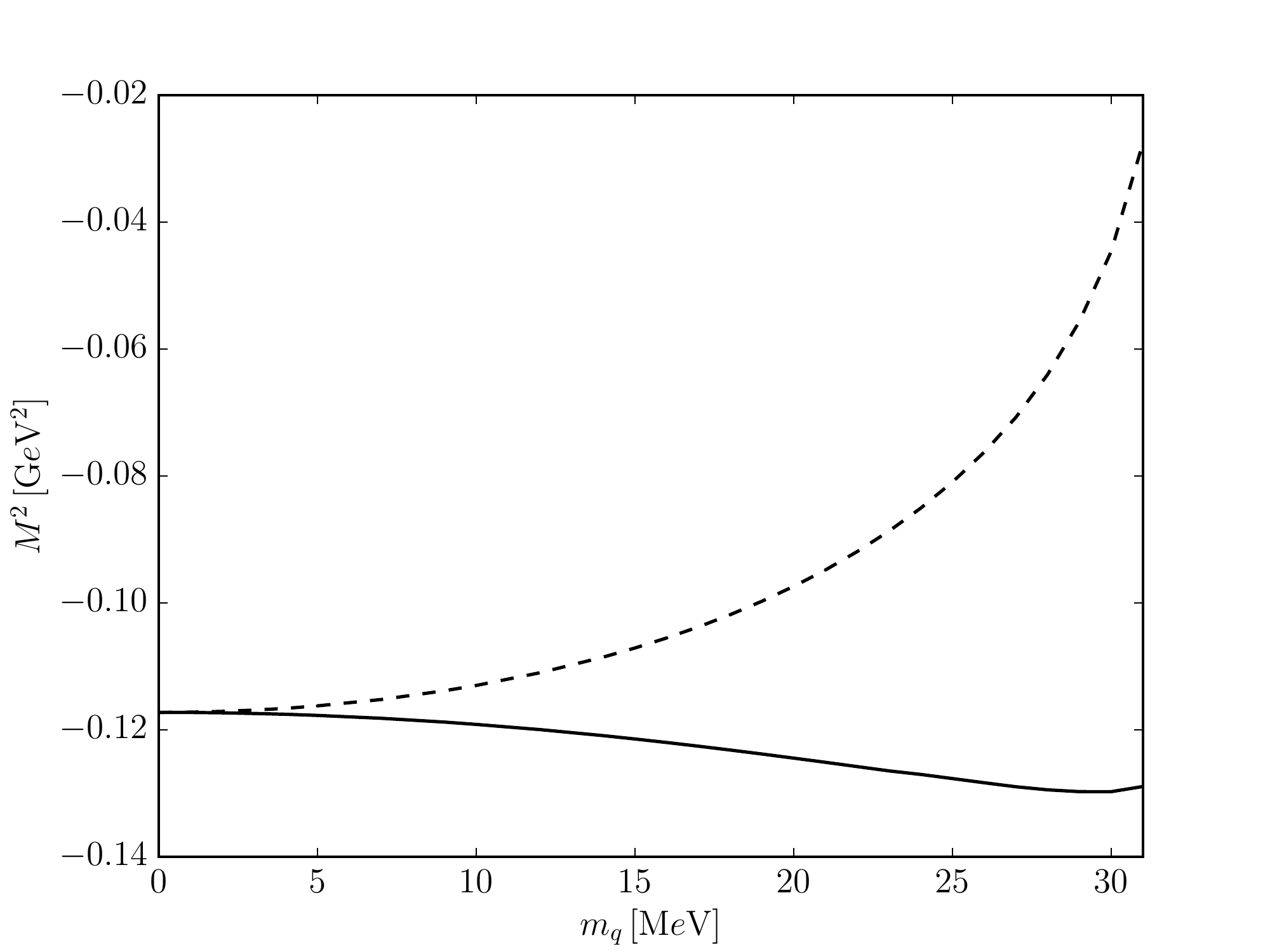}
\caption[Bound state mass $M^2$ in WW phase up to the critical current-quark mass.]{
Bound state masses in \gls{WW} phase for equal quarks in the pseudoscalar (solid) and scalar (dashed) channel up to the critical current-quark mass $m_q^{\mathrm{cr}}=31$\,M$e$V, $\omega = 0.5$\,G$e$V and $D=16.0$\,G$e$V$^2$.}
\label{fig:WW_M}%
\end{figure}

Having \gls{WW} solutions for the quark \gls{DS} equation beyond the chiral limit at disposal, one is able to study a scenario only with explicit symmetry breaking in the scope of a coupled \gls{DS}--\gls{BS} approach.
Such a scenario reveals the amount of mass splitting in the parity doublet caused by finite quark masses, the effect of explicit symmetry breaking on order parameters and allows for qualitative discussions related to \gls{DCSB}, its restoration and the relation to confinement.
For the scalar and pseudoscalar channel the masses are shown in Fig.\,\ref{fig:WW_M}.
They are spacelike over the whole current-quark mass region.
Moreover, the pseudoscalar squared bound state mass $M^2$ is even decreasing with increasing current-quark mass.
While, the modulus of the pion mass scarcely changes by $20$\,M$e$V for a change of the current-quark mass of $30$\,M$e$V and stabilizes towards $m_q^\mathrm{cr}$, the (imaginary) scalar mass decreases by more than $120$\,M$e$V with still increasing slope modulus.
As expected, the chiral limit behaviour in \gls{WW} phase qualitatively differs significantly from the limit in \gls{NG} phase.
In \gls{NG} phase a strong current-quark mass dependence of the scalar and, even more, the pseudoscalar bound state mass is observed, whereas the slope of the bound state mass curve in \gls{WW} phase approaches zero in the chiral limit.
Evaluating the formal splitting of scalar and pseudoscalar mesons at a current-quark mass of $m_q = 5$\,M$e$V yields $\vert \Delta M \vert = 1.6$\,M$e$V, which is tiny as compared to the splitting of chirality partners due to \gls{DCSB} (roughly $\vert \Delta M \vert \approx 350 \ldots 450$\,M$e$V).
However, it is of the order of the mass splitting in the isospin multiplet.

\begin{figure}%
\centering
\includegraphics[width=.45\textwidth]{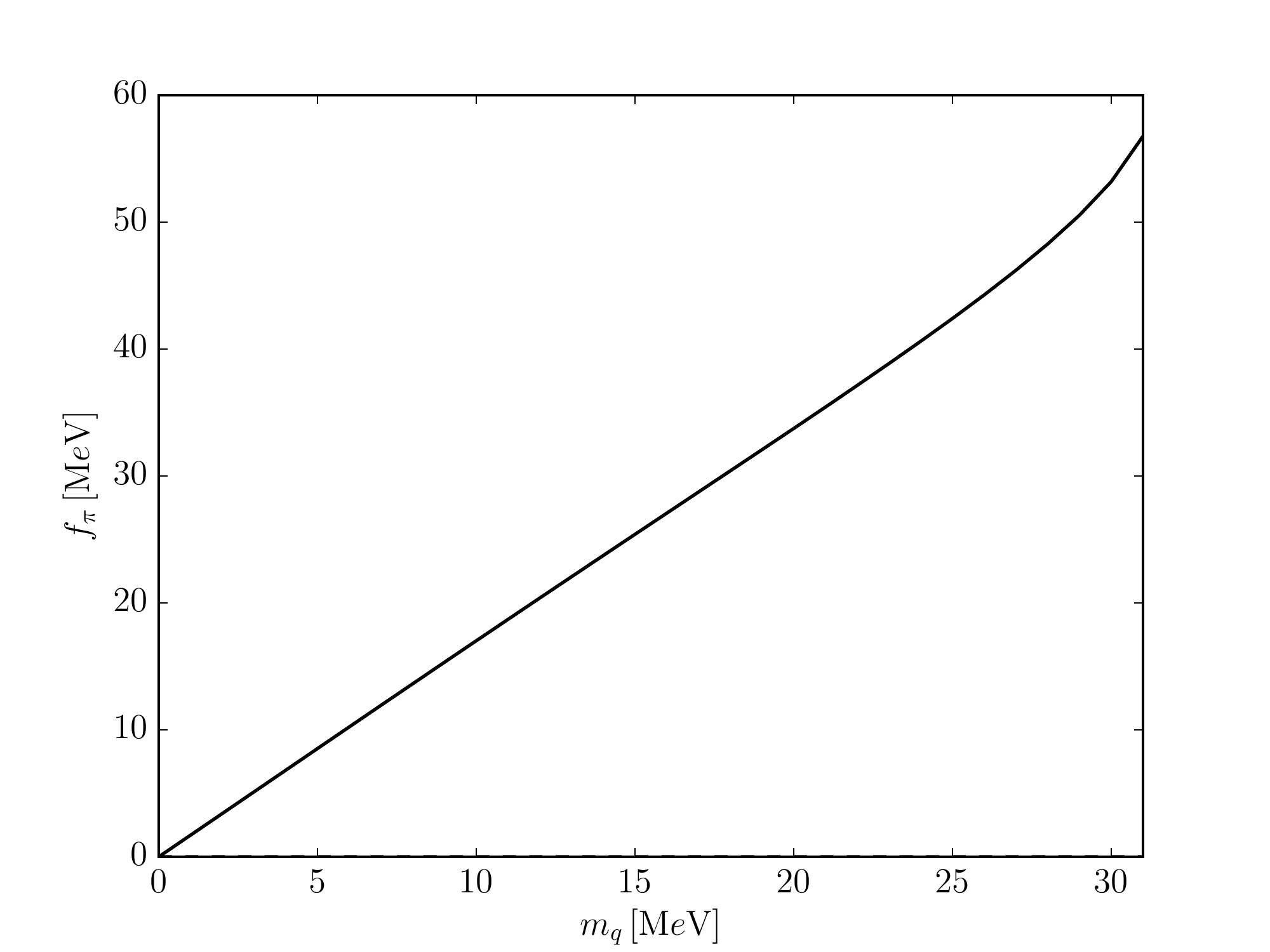}
\caption[Leptonic decay constant in WW phase up to the critical current-quark mass.]{
Leptonic decay constant in \gls{WW} phase for equal quarks in the pseudoscalar channel up to the critical current-quark mass $m_q^{\mathrm{cr}}=31$\,M$e$V, $\omega = 0.5$\,G$e$V and $D=16.0$\,G$e$V$^2$.
The scalar leptonic decay constant is numerically zero, $f_\upsigma \lessapprox 10^{-14}$\,G$e$V, for all current-quark masses as in \gls{NG} phase \cite{Bhagwat:2006py}.}
\label{fig:WW_f}%
\end{figure}

The leptonic decay constant, $f_\uppi$, in \gls{WW} phase is depicted in Fig.\,\ref{fig:WW_f}.
Apparently, $f_\uppi$ rises linearly with $m_q$.
Hence, the current-quark mass dependence in \gls{WW} phase qualitatively differs from the \gls{NG} phase \cite{Holl:2004fr}.
Interestingly, the absolute change of $f_\uppi$ of $\approx 50$\,M$e$V is larger than the corresponding change in \gls{NG} phase.
For $m_q = 5$\,M$e$V, $f_\uppi \approx 8.5$\,M$e$V, which is $\approx 9\,\%$ of its \gls{NG} phase value.
Note, that over the whole current-quark mass region employed within this study, deviations from the generalized \gls{GMOR}, Eq.\ \eqref{eq:GMOR}, are of the order $10^{-3}$ or below.

\begin{figure}%
\centering
\includegraphics[width=.45\textwidth]{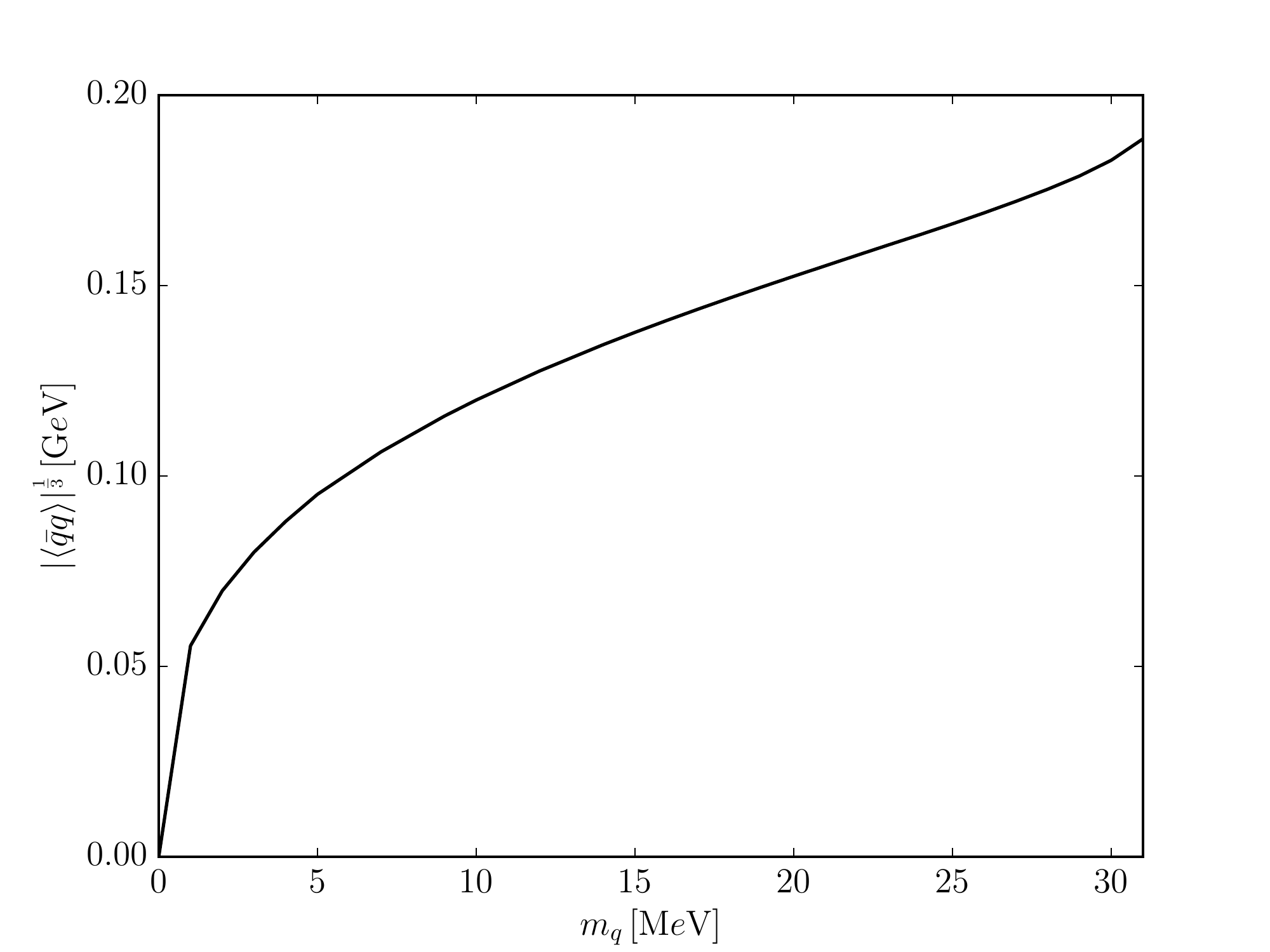}
\caption[Chiral condensate in WW phase up to the critical current-quark mass.]{
Chiral condensate in \gls{WW} phase up to the critical current-quark mass $m_q^{\mathrm{cr}}=31$\,M$e$V, $\omega = 0.5$\,G$e$V and $D=16.0$\,G$e$V$^2$.}
\label{fig:WW_qbarq}%
\end{figure}

Finally, in Fig.\,\ref{fig:WW_qbarq} the chiral condensate in \gls{WW} phase is depicted beyond the chiral limit, cf.\ Eq.\,\eqref{eq:qq}, and reveals the impact of explicit chiral symmetry breaking only.
Similar to $f_\uppi$, it rises linearly with the current-quark mass up to $m_q\approx 25$\,M$e$V.
Explicit chiral symmetry breaking by current-quark masses of $m_q\approx 30$\,M$e$V mimics roughly $40-50\,\%$ of the \gls{NG} phase values for leptonic decay constant and chiral condensate.
For $m_q = 5$\,M$e$V, $\left|\langle \ld \bar q q \rd \rangle\right| \approx (95\,\mathrm{M}e\mathrm{V})^3$, which is less than $6\,\%$ of its \gls{NG} phase value $\left|\langle \ld \bar q q \rd \rangle\right| \approx (256\,\mathrm{M}e\mathrm{V})^3$.

\section{Conclusions and outlook}
\label{sct:conclucions}

The bound state masses of pseudoscalar and scalar mesons within the Poincar\'e covariant, symmetry preserving, \gls{RL} truncated \gls{DS}--\gls{BS} approach with a model closely related to the phenomenologically successful \gls{MT} interaction have been studied in the chiral limit and beyond.
The validity of the \gls{AVWTI} in \gls{WW} phase has been confirmed.
It has been found that the spin-0 states disappear from the timelike spectrum and become spacelike over the whole accessible current-quark mass range.
Furthermore, the mass splitting between chirality partners due to explicit current-quark masses in \gls{WW} phase has been quantified to be of the same order as the experimental splitting in the \gls{NG} phase isospin multiplet.
Finally, the current-quark mass dependence of the pion leptonic decay constant and the chiral condensate in \gls{WW} phase has been revealed.
A strong linear dependence of both on the current-quark mass $m_q$ has been found, driving $f_\uppi$ and $\langle \ld \bar q q \rd \rangle$ in \gls{WW} phase at the critical current-quark mass $m^{\mathrm{cr}}_q$ above $40-50\,\%$ of their \gls{NG} phase values at $m_q = 5$\,M$e$V.
Whereas the difference between the \gls{NG} and \gls{WW} phase propagator is qualitatively clearly visible in the quark mass function, $M(p^2)$ (or $B(p^2)$), the difference in $\langle \ld \bar q q \rd \rangle$ or the observable quantity $f_\uppi$ is less pronounced when allowing for larger current-quark mass in \gls{WW} phase.

In view of the investigation of \cite{Hilger:2010cn} the extension of the above presented analysis to excited states, where the effects of \gls{DCSB} are suppressed, the spin-1 channel, and beyond, is in order.
It is not expected that higher spin states have spacelike solutions, either.
Similarly, the presented approach allows for the evaluation of decay properties on the same footing.
Furthermore, based on a phenomenological interaction which successfully describes the hadronic spectrum, the solutions of the \gls{DS} equation in \gls{WW} phase may be used to determine condensates \cite{Nguyen:2010yh,Nguyen:2010yj,Nguyen:2009if}, in particular the symmetric four-quark condensate which dominates the chirally symmetric \glspl{QSR} for the $\uprho$ meson.
This provides a reliable implicit relation between changes of chirally symmetric and chirally odd condensates, which may be employed in a \gls{QSR} calculation as in \cite{Hilger:2010cn}, and allows for a more sophisticated analysis.

\begin{acknowledgments}
The author acknowledges discussions with A.\ Krassnigg, M.\,Pak and G.\,Eichmann.
This work was supported by the Austrian Science Fund (FWF) under Grant No.\ P25121-N27.
\end{acknowledgments}

\end{document}